%% file: main.tex
\documentclass[conference]{IEEEtran}
\IEEEoverridecommandlockouts
\usepackage{cite}
\usepackage{amsmath,amssymb,amsfonts}
\usepackage{algorithmic}
\usepackage{graphicx}
\usepackage{xcolor}
\usepackage{textcomp}
\usepackage[T1]{fontenc}
\def\BibTeX{{\rm B\kern-.05em{\sc i\kern-.025em b}\kern-.08em
    T\kern-.1667em\lower.7ex\hbox{E}\kern-.125emX}}
    
\newcommand{\commscope}{Comm|Scope}

\usepackage{booktabs}
\usepackage{makecell}
\usepackage{subcaption}
\usepackage{multirow}

\usepackage{comment}
\usepackage{caption}

\usepackage{listings}
\definecolor{PaleBlue}{RGB}{230, 245, 255}
\definecolor{codegreen}{rgb}{0,0.6,0}
\definecolor{codegray}{rgb}{0.5,0.5,0.5}
\definecolor{codepurple}{rgb}{0.58,0,0.82}
\definecolor{backcolour}{rgb}{1,1,1}
\lstdefinestyle{mystyle}{
    backgroundcolor=\color{backcolour},   
    commentstyle=\color{codegreen},
    keywordstyle=\color{magenta},
    numberstyle=\tiny\color{codegray},
    stringstyle=\color{codepurple},
    escapebegin=\color{codegreen},
    basicstyle=\ttfamily\scriptsize,
    breakatwhitespace=false,         
    breaklines=true,                 
    captionpos=b,                    
    keepspaces=true,                 
    numbers=none,
    frame=single,
    numbersep=5pt,                  
    showspaces=false,                
    showstringspaces=false,
    showtabs=false,                  
    tabsize=2
}
\colorlet{backgroundcol}{cyan!10!white}

\makeatletter
\newcommand{\newlineauthors}{%
  \end{@IEEEauthorhalign}\hfill\mbox{}\par
  \mbox{}\hfill\begin{@IEEEauthorhalign}
}
\makeatother
\usepackage{hyperref}

\begin{document}

\title{Understanding Data Movement in AMD Multi-GPU Systems with Infinity Fabric}

\author{\IEEEauthorblockN{Gabin Schieffer}
\IEEEauthorblockA{\textit{KTH Royal Institute of Technology}\\
Stockholm, Sweden \\
gabins@kth.se}
\and
\IEEEauthorblockN{Ruimin Shi}
\IEEEauthorblockA{\textit{KTH Royal Institute of Technology}\\
Stockholm, Sweden \\
ruimins@kth.se}
\and
\IEEEauthorblockN{Stefano Markidis}
\IEEEauthorblockA{\textit{KTH Royal Institute of Technology}\\
Stockholm, Sweden \\
markidis@kth.se}
\newlineauthors
\IEEEauthorblockN{Andreas Herten}
\IEEEauthorblockA{\textit{Jülich Supercomputing Centre}\\
Jülich, Germany \\
a.herten@fz-juelich.de}
\and
\IEEEauthorblockN{Jennifer Faj}
\IEEEauthorblockA{\textit{KTH Royal Institute of Technology}\\
Stockholm, Sweden \\
faj@kth.se}
\and
\IEEEauthorblockN{Ivy Peng}
\IEEEauthorblockA{\textit{KTH Royal Institute of Technology}\\
Stockholm, Sweden \\
ivybopeng@kth.se}
}

\maketitle

\begin{abstract}

Modern GPU systems are constantly evolving to meet the needs of computing-intensive applications in scientific and machine learning domains. However, there is typically a gap between the hardware capacity and the achievable application performance. This work aims to provide a better understanding of the Infinity Fabric interconnects on AMD GPUs and CPUs. We propose a test and evaluation methodology for characterizing the performance of data movements on multi-GPU systems, stressing different communication options on AMD MI250X GPUs, including point-to-point and collective communication, and memory allocation strategies between GPUs, as well as the host CPU. In a single-node setup with four GPUs, we show that direct peer-to-peer memory accesses between GPUs and utilization of the RCCL library outperform MPI-based solutions in terms of memory/communication latency and bandwidth. Our test and evaluation method serves as a base for validating memory and communication strategies on a system and improving applications on AMD multi-GPU computing systems.
\end{abstract}

\begin{IEEEkeywords}
AMD MI250X GPU, Multi-GPU Programming, AMD MI250X Memory System Performance
\end{IEEEkeywords}

\input{10_intro.tex}

\input{20_background.tex}
\input{30_method.tex}

\input{40_cpugpu}

\input{50_p2p_explicit}
\input{51_p2p_stream}

\input{53_p2p_MPI}

\input{54_collectives}

\input{60_related.tex}

\input{70_conclusion.tex}

\section*{Acknowledgment}
This work is funded by the European Union. This work has received funding from the European High Performance Computing Joint Undertaking (JU) and Sweden, Finland, Germany, Greece, France, Slovenia, Spain, and the Czech Republic under grant agreement No.~101093261. The computations were enabled by resources provided by the National Academic Infrastructure for Supercomputing in Sweden (NAISS), partially funded by the Swedish Research Council through grant agreement no. 2022-06725. This research is supported by the Swedish Research Council (no. 2022.03062).

\bibliographystyle{IEEEtran}
\bibliography{main}

\end{document}

%% file: 10_intro.tex
\section{Introduction}

\begin{figure}[t]
    \centering
    \includegraphics[width=\linewidth]{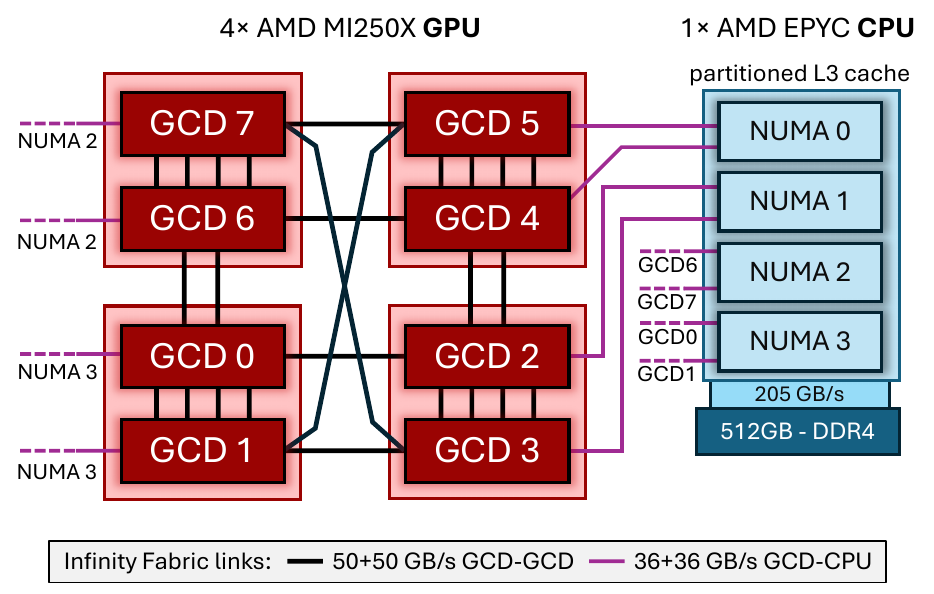}
    
    \caption{Overview of a multi-GPU compute node, totaling eight~GCDs, distributed onto four physical MI250X GPUs, coupled with a single-socket AMD 3\textsuperscript{rd}~generation EPYC CPU. Adapted from~\cite{frontier-user}.}

    \label{fig:arch}
\end{figure}

Multi-GPU HPC nodes have become omnipresent in large-scale supercomputers to support a variety of accelerated scientific workloads, ranging from weather forecast~\cite{geenen2024digital}, computational fluid dynamics~\cite{karp2023large,jansson2023exploring}, molecular dynamics~\cite{andersson2022breaking}, plasma simulation~\cite{williams2024understanding}, and quantum computer simulators~\cite{markidis2023enabling}. Currently, large-scale HPC clusters exhibit computing nodes with multiple GPUs on the same node, interconnected via a high-performance interconnect or through PCIe. While multi-GPU supercomputers were initially dominated by Nvidia, today a large number of supercomputers also rely on AMD GPUs to accelerate scientific workloads~\cite{loh2023research}. For instance, ORNL Frontier features four AMD Instinct MI250X GPUs per computing node, each GPU comprises two Graphics Compute Dies~(GCD). From the user perspective, one computing node appears as an eight-GPU node, where each GCD is seen as a GPU. On this system, CPU and GPUs are connected via the in-package Infinity Fabric high-performance interconnect, similar to Nvidia's NVLink. This interconnect creates multiple paths for data movements between all processors in the system, CPU and GPUs, and can be utilized using various interfaces. In this mesh, the various links exhibit three different bandwidth levels, which significantly complexifies the design of high-performance applications, that wish to efficiently leverage the hardware capabilities of this system. Other supercomputers, such as LUMI, also use this unique node architecture.

Historically, the performance of GPU-accelerated systems has been limited by the available latency and bandwidth between the host and GPU. Several technological improvements, such as specialized interconnects to enhance bandwidth, or overlapping techniques have been designed to address this issue. Today, with multiple GPUs per node, the problem of data movement performance across several devices, all serving the same application code, becomes even more serious. For this reason, it is imperative to understand the performance benefits and bottlenecks of data movement and characterize its performance between CPU and GPUs, and across different GPUs on the same node. Such understanding can help scientists in designing data placement and migration techniques for application and runtime systems.

The overall goal of this paper is to test and characterize the data movement performance on AMD multi-GPU nodes, where CPU-GPU and GPU-GPU are interconnected with Infinity Fabric. This work is based on a compute node with eight GCDs, grouped onto four physical AMD MI250X GPUs, which is similar to the topology used in Frontier, the No.1 supercomputer on the Top500 list\footnote{\url{https://top500.org/lists/top500/2024/06/} (Accessed on August 2024)}. Our testing methodology consists of three steps, covering three essential usage scenarios of the Infinity Fabric interconnect. First we characterize performance of CPU-GPU data movements, comparing available programming interfaces and memory allocations. We then focus on the GPU-GPU Infinity Fabric interconnect, in the context of peer-to-peer data movements. Finally, we provide an evaluation of high-level multi-GPU collectives, which are widely used in both HPC and AI applications, in the form of MPI collectives and RCCL collectives. To enable reproducibility, we provide the code and benchmark scripts\footnote{\url{https://github.com/KTH-ScaLab/multi-gpu-comm}}.

The main contributions of this paper are summarized as follows:
\begin{itemize}
    \item We survey the various point-to-point data movement options on a compute node composed of an AMD CPU and four MI250X GPUs with eight GCDs interconnected via Infinity Fabric.
    \item We evaluate the impact of memory allocation strategies on the performance of data movement between AMD CPU and GPU. 
    \item We characterize the performance of collective communication and compare MPI and RCCL libraries on multi-GPU nodes with topology representative of the Frontier/LUMI supercomputer.
    \item We identify the importance of optimal routing in the complex Infinity Fabric topology.
\end{itemize}

%% file: 20_background.tex
\section{Background}
\label{sec:bg}
In this work, we focus on an HPC multi-GPU node, featuring four AMD MI250X GPUs, and a third generation AMD EPYC CPU, along with Infinity Fabric interconnect for CPU-GPU and inter-GPU communication. Figure~\ref{fig:arch} illustrates the inter-connectivity of such multi-GPU node. An important and unique particularity of this system is that each MI250X GPU is built up of two Graphics Compute Dies (GCDs) with 64~GB HBM2e memory per GCD, offering a peak bandwidth of 1.6~TB/s. Each GCD has an 8~MB L2 cache shared between all compute units, each compute unit has 16~KB of L1 vector cache and 16~KB of L1 scalar cache, shared by each pairs of two compute units. From a user's perspective, each GCD behaves as a single GPU, as a GCD features its own compute units and its own physical memory. The CPU in this node is a AMD third generation EPYC processor, of Zen~3 micro-architecture, specifically a 64-core AMD 7A53 CPU. The CPU is attached to 512~GB of DDR4 memory, which is divided into four NUMA domains, with each pair of two GCDs (one physical MI250X GPU) being connected to exactly one NUMA domain, as depicted in Figure~\ref{fig:arch}. Another important particularity of this system is the Infinity Fabric interconnect, offering high-performance CPU-GPU, along with GPU-GPU communication abilities. This node characteristics and topology are similar as for computing nodes of the Frontier~\cite{frontier} and LUMI~\cite{lumi} supercomputers.

\subsection{Infinity Fabric Interconnect}
The GCDs within one and across GPUs are connected through Infinity Fabric links. Figure~\ref{fig:arch} presents the complete node topology. The Infinity Fabric between the different GCDs are implemented as a single, dual or quad connections of 50+50~GB/s\footnote{in this paper, $1~\text{GB/s}=10^9~\text{bytes/s}$} bidirectional bandwidth per link. GCDs residing on the same physical GPU are connected through four links, resulting in a total bidirectional bandwidth of 400~GB/s. Taking GCD0 as an example, it is also directly connected through a dual link to GCD6 (i.e., 200~GB/s bidirectional) and through a single link to GCD2 (i.e., 100~GB/s bidirectional). All other GCDs can be reached through two hops from GCD0. Underlyingly, the Infinity Fabric interconnect implements the xGMI protocol; with various numbers of xGMI links interconnecting GCDs. Each link operates on 16 bits per transaction, with a transaction rate of 25~GT/s, that is, a 50~GB/s peak bandwidth per link per direction~\cite{mi250_microarch}. Furthermore, each GCD is connected to the host CPU of the system, through a single Infinity Fabric link, with a theoretical peak bandwidth of 36~GB/s (72~GB/s bidirectional). The Infinity Fabric interconnect supports zero-copy memory access, where any processor in the system, CPU or GPU, can access each other's physical memory, directly over the interconnect, without the need to maintain a local copy.

\subsection{HIP Programming Model}
AMD GPUs are predominantly programmed using the HIP programming model, which is a C++ based runtime API and kernel language, similar to the established CUDA programming environment for Nvidia GPUs. Underlyingly, at runtime, the HIP runtime interacts with the HSA runtime, which in turns communicates with the AMD kernel driver (ROCk). AMD's ROCm platform provides compilers and development tools to program GPUs with HIP as well as HIP versions of common HPC libraries, such as rocBLAS. It also offers a command line tool to translate CUDA code into HIP code (\emph{hipify}), which is used in the course of this work.

\subsection{Memory Management}
The topology of a multi-GPU node makes memory allocation and data transfer a complex task from the user's perspective, as the physical memory in this system is distributed across eight GPUs and a CPU, where it is further divided into four NUMA domains. To abstract this complexity, AMD provides several APIs to allocate memory and perform inter-GPU communication, with various levels of abstraction and granularity, those APIs are similar to the ones found in the CUDA programming model.

Unified Memory groups into a single virtual memory space the physical memory of all processors in the system. This means that any processor, CPU or GPU, can access other processors' physical memory using a single virtual address. Unified Memory can be allocated through managed memory, using \texttt{hipMallocManaged}. In addition, other type of memory allocation can be \textit{mapped} into the GPU's virtual address space, either automatically by the runtime, at allocation time, or explicitely using HIP APIs. Mapping memory allows the GPU to access memory outside its physical memory. 

Memory allocations can be configured as \textit{coherent}. When a GPU modifies CPU memory that is marked as coherent, changes are immediately reflected on the CPU-side. On MI250X, to achieve this effect, GPU-side caching is disabled for coherent memory. Therefore, each access to data located in remote coherent memory generates traffic over the CPU-GPU interconnect. While the use of coherent memory is generally detrimental to performance, it simplifies programming for application with complex access patterns, for example with co-running CPU and GPU kernels operating on the same data. Note that on more recent systems, such as AMD MI300A, the no-caching restriction can be lifted thanks to the introduction of cache-coherent interconnects. In HIP, by default, host-pinned memory is marked as coherent. 

When a GPU kernel accesses memory that is neither located in the GPU's physical memory, nor mapped into GPU virtual memory, a page fault is triggered. MI250X systems have the ability to resolve the page fault, and retry the memory access. This feature, referred to as \textit{XNACK}, can be enabled by setting the environment variable \texttt{HSA\_XNACK=1}. Additionally, users must ensure that GPU kernels are built to match the system's XNACK configuration (enabled/disabled). %

Table~\ref{tab:hip_alloc} summarizes a list of memory allocation APIs in HIP. The second column further details how memory movements are performed. Three types of data movement are listed, namely explicit, implicit, and zero-copy. \textit{Explicit} data movement refers to the use of \texttt{hipMemcpy}, where the user is responsible for transferring data. \textit{Zero-copy} indicates that data is accessed over Infinity Fabric Interconnect. Finally, \textit{implicit} data movement indicates that memory is automatically migrated as it is accessed. Migration are performed as the page granularity, where an entire page is migrated, independent of the size of the data being accessed. This is the behavior for \texttt{hipMallocManaged} memory, when XNACK is enabled.

\begin{table*}[ht]
    \centering
    \caption{Memory allocation methods in HIP for CPU-side allocation, and strategy for CPU-GPU data movement.}
    \begin{tabular}{ccccc}
        \toprule
         Memory & Data Movement & Coherence & API: Allocation & API: Data movement \\
         \midrule\midrule
         Pinned & explicit & no & {hipHostMalloc(flag=hipHostMallocNonCoherent)} & hipMemcpy(Async) \\
         Pageable & explicit & no & {malloc} & {hipMemcpy} \\
         \midrule
         Pinned & zero-copy & yes & hipHostMalloc([flag=hipHostMallocCoherent]) & / \\
         \midrule
         Unified	& zero-copy & yes & hipMallocManaged(); HSA\_XNACK=0 & / \\ 
         Unified	& implicit & yes & hipMallocManaged(); HSA\_XNACK=1 & / \\
         \bottomrule
    \end{tabular}
    \label{tab:hip_alloc}
\end{table*}

%% file: 30_method.tex
\section{Testing Methodology}
\label{sec:method}

\begin{table*}[t]
    \caption{Description of the evaluated memory types, corresponding benchmarks and programming interfaces}
    \centering
    \label{tab:experiments}
\begin{tabular}{ccccc}
\toprule
Link & Category   & Benchmark     & Allocation & Data movement \\
\midrule\midrule
/ & Local GPU memory & STREAM (Copy) & \texttt{hipMalloc} & local access (GPU kernel) \\ \midrule
\multirow{5}[3]{*}{\rotatebox[origin=c]{90}{CPU-GPU}} & \multirow{5}[3]{*}{CPU-GPU} 
& \multirow{4}{*}{\commscope}  & pageable (\texttt{malloc})          & \texttt{hipMemcpy} \\
                             &&& pinned (\texttt{hipHostMalloc})     & \texttt{hipMemcpy} \\
                             &&& managed (\texttt{hipMallocManaged}) & zero-copy (GPU kernel) \\
                             &&& managed (\texttt{hipMallocManaged}) & page migration (XNACK) \\ \cmidrule{3-5}
&& STREAM (copy)               & pinned (\texttt{hipHostMalloc}) & zero-copy (GPU kernel) \\

\midrule
\multirow{6}[8]{*}{\rotatebox[origin=c]{90}{GPU-GPU}} & \multirow{3}[3]{*}{GPU peer-to-peer} & \commscope & \texttt{hipMalloc} & \texttt{hipMemcpyPeer} \\ \cmidrule{3-5}
&& p2pBandwidthLatencyTest & \texttt{hipMalloc} & \texttt{hipMemcpyPeer}\\ \cmidrule{3-5}
&& STREAM (copy) & \texttt{hipMalloc} & zero-copy (GPU kernel)  \\ \cmidrule{2-5}
& MPI GPU point-to-point  & OSU micro-benchmarks    & \texttt{hipMalloc} & \texttt{MPI\_ISend}, \texttt{MPI\_Recv} \\ \cmidrule{2-5}
& \multirow{2}[2]{*}{MPI GPU Collectives}  & OSU micro-benchmarks    & \texttt{hipMalloc} & MPI collectives \\ \cmidrule{3-5}
&& RCCL-tests & \texttt{hipMalloc} & RCCL collectives \\
\bottomrule
\end{tabular}
\end{table*}

In this work, we employ a testing methodology that quantifies and validates the achievable performance of data movements over Infinity Fabric between different computing pairs, including CPU-GPU and GPU-GPU. The detail of tools and benchmarks used in this paper is presented in Table~\ref{tab:experiments}. In general, we evaluate the bandwidth of the data movements, for various transfer sizes, along with the latency for GPU-GPU communication.

For CPU-GPU data movements, we first identify a baseline peak achievable bandwidth for different memory allocation and memory access interfaces. We use for this purpose the host-to-device test cases from the \commscope~\cite{pearson2019} microbenchmarks. \commscope{ }is a set of microbenchmarks, focusing on data movements on multi-GPU multi-CPU systems. It provides test cases for various data placement scenarios and data movement interfaces. We further use a variant of the STREAM benchmark to evaluate the performance of direct memory access to CPU memory from GPU kernels. Each of the eight GCDs on a MI250X system is presented to the user as a GPU that can be programmed independently of others. To evaluate this aspect, we scale our simple CPU-GPU STREAM benchmark in parallel from one GCD up to the eight GCDs on the system, we also evaluate the impact of placement strategy on overall total CPU-GPU bandwidth.

To evaluate the performance of the GPU-GPU Infinity Fabric interconnect, we use a similar approach as for CPU-GPU, where we both evaluate the performance of explicit data movements, relying on \texttt{hipMemcpyPeer}, and direct memory access with GPU kernels. In this evaluation, we compare the physical topology, composed of various tiers of GCD-GCD Infinity Fabric links, as illustrated in Figure~\ref{fig:arch}, with the achieved performance. Those experiments are summarized in Table~\ref{tab:experiments}. In this set of tests, in addition to the STREAM benchmark and \commscope, we use a HIP-ported version of p2pBandwidthLatencyTest, a benchmark provided by Nvidia to measure performance of peer-to-peer data movements~\cite{cudasamples}. By comparing the obtained performance from using these options of APIs with the theoretical peak performance, we are able to validate and assess the performance of data movements.

As HPC users are highly reliant on MPI to implement inter-process communication in their applications, it is crucial to evaluate the performance of MPI routines in the context of multi-GPUs applications. Therefore, we complete our GPU peer-to-peer analysis with a point-to-point bandwidth test from the OSU micro-benchmarks suite~\cite{omb}, which relies on MPI for communication. In this experiment, we aim at understanding the underlying transfer interface used by MPI communication, along with the ability to leverage direct GPU-GPU data movements over Infinity Fabric interconnect.

Besides point-to-point communication, performance of multi-GPU collective communication is highly-relevant, both for HPC and AI applications. While MPI is a viable option, specialized libraries that implement multi-GPU collectives are widely used, especially in AI applications. Nvidia Collective Communication Library (NCCL) and AMD's RCCL library are two widely used GPU collectives libraries. In this paper, we evaluate the performance of GPU collectives both for MPI and RCCL. We use for this purpose the OSU collective micro-benchmarks and the RCCL-tests library. This choice of benchmark is summarized in Table~\ref{tab:experiments}. We measure the latency of five collectives, namely Reduce, Broadcast, AllReduce, ReduceScatter and AllGather, covering three communication patterns, namely all-to-one, one-to-all, and all-to-all. We compare our results with a simple lowest theoretical bound, based on latency measured with previous GPU-GPU latency test.

For our experiments, we use ROCm~5.7.0, which includes the HIP runtime and the RCCL library 2.17.1. As our toolchain, we use the compiler from the Cray Programming Environment~23.12, along with the \texttt{hipcc} compiler, part of the ROCm installation; both compilers are based on LLVM/Clang~17. As the MPI implementation, we use Cray-MPICH~8.1.28 available on our system, which we configure to be GPU-aware, by enabling the environment variable \texttt{MPICH\_GPU\_SUPPORT\_ENABLED=1}. We use version 7.4 of the OSU Micro-benchmarks.

%% file: 40_cpugpu.tex
\section{CPU-GPU Communication}
\label{sec:cpugpu}
\begin{figure}[ht]
    \centering
    \includegraphics[width=\linewidth]{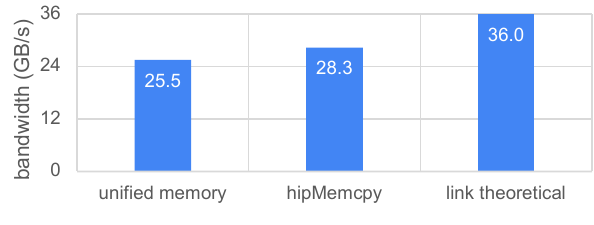}
    \caption{Peak achieved host-to-device bandwidth in our experiments, for direct GPU access to CPU memory with unified memory, and explicit data movements with hipMemcpy}
    \label{fig:cpugpu_summary_h2d}
\end{figure}

Data movement between CPU and GPU is performed through the CPU-GPU Infinity Fabric links. A single Infinity Fabric link has a  theoretical bandwidth of 36~GB/s per direction (72~GB/s bidirectional), and each GCD is connected to the CPU with exactly one of these links. For reference, the memory latency of the DDR memory of our AMD EPYC CPU is 96~ns; the CPU memory bandwidth is 204.8~GB/s. Figure~\ref{fig:cpugpu_summary_h2d} summarizes our results for CPU to GPU data movements, obtained with \commscope, using either unified memory or \texttt{hipMemcpy}, which we detail in this section.

\subsection{Peak Achievable Bandwidth}
\begin{figure}[bt]
    \centering
    \includegraphics[width=\linewidth]{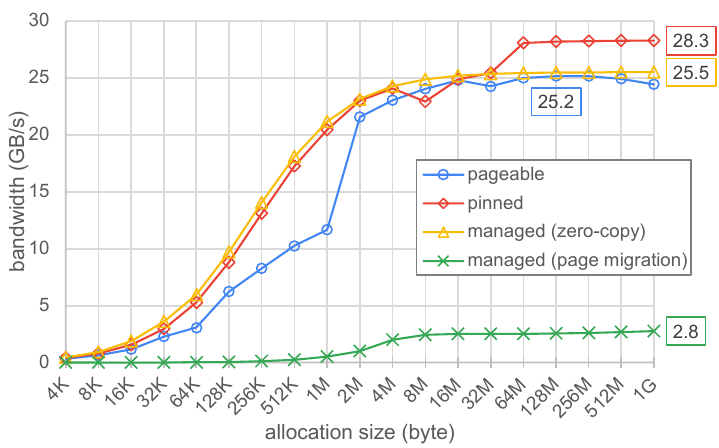}
    \caption{Host-to-device memory bandwidth at increased data transfer sizes, measured with \commscope. The maximum for each interface is indicated in boxes.}
    \label{fig:comm_scope_h2d}
\end{figure}

We use \commscope~\cite{pearson2019} to evaluate the peak achievable CPU-GPU bandwidth using both unified memory, and explicit data movements, in a similar fashion as proposed in~\cite{pearson2023interconnect}. Figure~\ref{fig:comm_scope_h2d} presents the results in the host-to-device direction, for transfer sizes sweeping from $4\mskip3mu \text{KB}$ to $1\mskip3mu \text{GB}$. Explicit data movements are performed with hipMemcpy, either from pageable memory (allocated with \texttt{malloc}), or with host-pinned memory (allocated with \texttt{hipHostMalloc}). Implicit data movements are performed using managed memory either through zero-copy, where the GPU directly accesses CPU-located memory, over the Infinity Fabric link, or with page migration, where pages accessed from the GPUs are migrated by the runtime when needed

We achieve a maximum bandwidth of 28.3~GB/s, with explicit data transfer from pinned memory. Pageable memory exhibits varying results when increasing the transfer size. This is expected, as non-predictable paging operations might reduce performance. For implicit data transfers, managed memory with page migration only achieved 2.8~GB/s, while managed memory with zero-copy access achieves a highest bandwidth of 25.5~GB/s.

In addition, zero-copy managed memory approximate the behavior of pinned memory, up to 32~MB transfer size, after which pinned memory bandwidth is able to reach higher value than managed memory. This observation could be an impact of the 32~MB L3 GPU cache. These results shows that zero-copy memory can achieve high utilization of the CPU-GPU link, which makes it an entailing programming interface for users.

\subsection{GPU-Aware Memory Placement}
The CPU memory of the investigated MI250X compute node is divided into four NUMA nodes. Each node is directly attached to the two GCDs of a physical GPU; this NUMA-to-GCD mapping can be obtained using the \texttt{rocm-smi -{}-showtoponuma} command. On our testbed, this mapping is identical to the one on Frontier and LUMI supercomputers, which is depicted in Figure~\ref{fig:arch}.

From the user's perspective, this mapping does not need to be known. Indeed, by default the \texttt{hipHostMalloc} API allocates pinned memory on the NUMA node closest to the current active GPU, selected with e.g., \texttt{hipSetDevice}. In our experiments, we rely on this behavior to allocate memory on the correct NUMA node.

To override this behavior, user can instruct the runtime to follow the user's NUMA placement policy instead, by passing the \texttt{hipHostMallocNumaUser} flag to \texttt{hipHostMalloc}. Other approaches are possible to achieve the same goal, such as allocating memory on a NUMA node with \texttt{numa\_alloc\_on\_node}, and then pinning it using \texttt{hipHostRegister}. Using \commscope's NUMA to GPU benchmark, which enforces data placement on a chosen NUMA node, we were not able to identify any bandwidth degradation when performing a copy operation within a non-optimal combination of NUMA node/GCD. This can be explained by the much higher inter-NUMA bandwidth, compared to the bandwidth over the interconnect.

\subsection{Multi-GPU Bandwidth}

To evaluate the behavior of multi-GPU host-to-device transfer, we use the STREAM copy kernel, launching one kernel per GPU in the system, from one GPU to the eight available GPUs. Using a single-threaded program, for each GPU, we allocate two pinned buffers on the CPU-side, using \texttt{hipHostMalloc}. We then launch one STREAM copy kernel per GPU, and enforce a CPU-GPU synchronization after kernel execution, for each GPU. This allows to measure the total execution time of kernels for all GPUs, from which we can obtain the total bandwidth. Listing~\ref{lst:multigpu_stream} presents this approach. The bidirectional bandwidth is then obtained with $BW=N_\text{GPU} \cdot 2N / t$, with $t$ the elapsed time, and $N$ the number of bytes in one buffer. In our experiments, we use $N=8\mskip3mu\text{GB}$.

\begin{lstlisting}[language=C++,caption={Multi-GPU CPU-GPU STREAM benchmark},label=lst:multigpu_stream,mathescape=true]
// allocate host-pinned buffers
for(int i = 0; i < num_gpus; i++) {
    hipSetDevice(i);
    hipHostAlloc(&a[i], N);
    hipHostAlloc(&b[i], N);
    init_array<<<...>(a[i], N);
}
// launch one kernel per GPU
t0 = clock();
for(int i = 0; i < num_gpus; i++) {
    hipSetDevice(i);
    STREAM_Copy<<<...>>>(a[i], b[i], N); // b[i] $\leftarrow$ a[i]
}
for(int i = 0; i < num_gpus; i++) {
    hipSetDevice(i);
    hipDeviceSynchronize();
}
t1 = clock();
\end{lstlisting}

We execute this benchmark respectively on one, two, four, and eight GCDs. In this experiment, attention should be paid to correctly launch the benchmark so that it utilizes the chosen physical GPUs. This can be achieved using the system's job scheduler, for example Slurm's \texttt{-{}-gpu-bind} option. However, this solution might not be supported on all systems, therefore, to execute our benchmark, we allocate all GPUs in one node to our benchmark's process, and use the \texttt{HIP\_VISIBLE\_DEVICES} environment variables to restrict the GPUs that are effectively used.

\begin{figure}[ht]
    \centering
    \includegraphics[width=\linewidth]{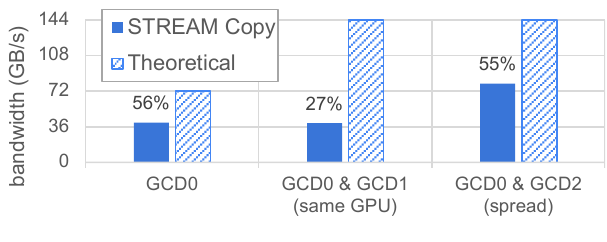}
    \caption{Total bidirectional CPU-GPU bandwidth, measured using STREAM copy kernels, parallelly-running on one or two GCDs. For the dual-GCDs cases, the two GCDs are either located on a single physical GPU (\textit{same GPU)}, or on two distinct physical GPUs (\textit{spread}). The achieved percentage of theoretical bandwidth is presented.}
    \label{fig:streamcpu_1and2}
\end{figure}

In a first experiment, we scale our benchmark from one GCD to two GCDs. For the dual-GCD execution, we evaluate two placement strategies: we chose either to execute on the two GCDs of the same physical GPU (\textit{same GPU}), or to spread the kernel launches across two GCDs belonging to distinct physical GPUs (\textit{spread}). Figure~\ref{fig:streamcpu_1and2} presents the achieved total bandwidth. We observe that only the \textit{spread} strategy scales correctly, as the bandwidth double from one to two GCDs in the \textit{spread} placement strategy. In contrast, using two GCDs of the same GPU does not provide a bandwidth improvement over single GCD. This could be an effect of each NUMA domain on the CPU handling two Infinity Fabric links~\cite{pearson2023interconnect}. Scenarios with lower transfer sizes, typically within CPU cache size range, could exhibit higher bandwidth over the Infinity Fabric links.

\begin{figure}[ht]
    \centering
    \includegraphics[width=\linewidth]{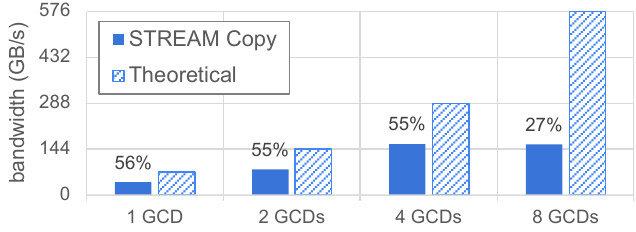}
    \caption{Total bidirectional CPU-GPU bandwidth, measured using STREAM copy kernels, parallelly-running on one to eight GCDs. The achieved percentage of theoretical bandwidth is presented.}
    \label{fig:streamcpu_1to8}
\end{figure}

Following the same approach, we repeat this experiment, scaling our benchmark from one to eight GCDs, using the \textit{spread} placement strategy. The aggregated bandwidth over all utilized links is presented in Figure~\ref{fig:streamcpu_1to8}, along with the theoretical bandwidth, and the achieved percentage of this bandwidth. We observe that scaling in the range 1-4 GCDs proportionally increases the bandwidth with the number of utilized GCDs. However, using eight GCDs does not improve the aggregated bandwidth, compared to four GCDs. This is expected, as we previously shown that using both GCDs of a single physical GPU -- which is the case in this experiment -- does not increase the measured bandwidth.

%% file: 50_p2p_explicit.tex
\section{Point-to-Point GPU Communication}
\label{sec:p2p}
In this section, we evaluate point-to-point communication between two GCDs. Two HIP interfaces, i.e., explicit data movements via the \texttt{hipMemcpyPeer} API, and unified memory, and GPU-aware MPI point-to-point can support users to perform such communication.

\subsection{Explicit Peer-to-Peer Data Movements}

\subsubsection{Latency}
\begin{figure*}[t]
    \begin{subfigure}{.333\textwidth}
        \centering
        \includegraphics[height=5.5cm]{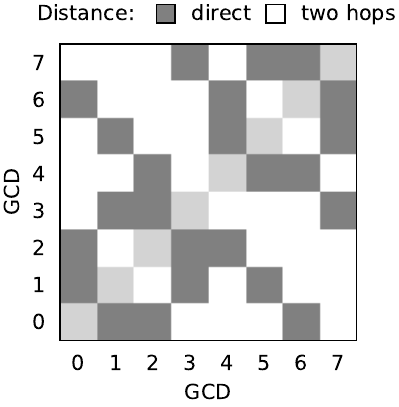}
        \caption{Length of shortest path}
        \label{fig:adjacency_mat}
    \end{subfigure} %
    \begin{subfigure}{.333\textwidth}
        \centering
        \includegraphics[height=5.5cm]{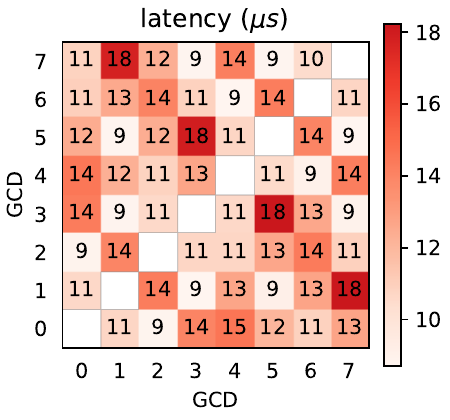}
        \caption{Latency}
        \label{fig:p2p_lat_mat}
    \end{subfigure} %
    \begin{subfigure}{.333\textwidth}
        \centering
        \includegraphics[height=5.5cm]{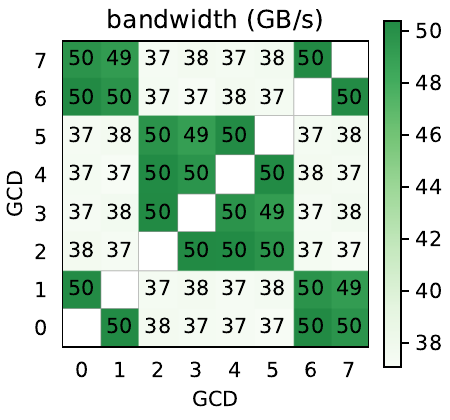}
        \caption{Bandwidth}
        \label{fig:p2p_bw_mat}
    \end{subfigure}
    \caption{Length of shortest path for each given GCD pairs (\subref{fig:adjacency_mat}), peer-to-peer GPU latency (\subref{fig:p2p_lat_mat}) and unidirectional bandwidth (\subref{fig:p2p_bw_mat}), measured with p2pBandwidthLatencyTest.}
\end{figure*}

We first quantify the latency of the \texttt{hipMemcpyPeer} operation using p2pBandwidthLatencyTest. As a reference for our analysis, Figure~\ref{fig:adjacency_mat} visualizes the length of the shortest path between all pairs of two given GCDs, in terms of number of hops. In this topology, the length of the shortest path never exceeds two hops. However, while such path is shortest in terms of number of hops, it does not maximize the bandwidth. For example, GCDs 1-7 are connected through a shortest path of two hops (1-3-7); however, the path maximizing the bandwidth is composed of three hops (1-0-6-7).

Using our HIPified version of p2pBandwidthLatencyTest, we measure the latency of peer-to-peer explicit data movements. For this purpose, we use the \texttt{hipMemcpyPeerAsync} API, with a transfer size of 16 bytes. The memory is allocated using \texttt{hipMalloc}, on both the source and destination GCDs. Memory is made available to peers using the \texttt{hipDeviceEnablePeerAccess} API. The latency is measured using the HIP Event API to time a \texttt{hipMemcpyPeerAsync} operation on the GPU-side. Each experiment is repeated 100 times. Results are presented as a matrix in Figure~\ref{fig:p2p_lat_mat}.

The measured latency varies within 8.7-18.2~$\mu s$. The latency measured between GCDs located on the same physical GPU is between 10.5-10.8~$\mu s$, which is not consistently lower that latency measured for other pairs of GCDs. Interestingly, the GCD pairs 0-2, 1-3, 1-5, 3-7, 4-6, 5-7 exhibit a latency below 10~$\mu s$. Comparing with the topology presented in Figure~\ref{fig:arch}, we observe that these pairs are exactly the ones which are interconnected by single Infinity Fabric link.

Furthermore, we observe four outliers, with latency values within 17.8-18.2$\mu s$, corresponding to the GCD pairs 1-7 and 5-3. We note that these two pairs are the only ones for which the bandwidth-maximizing path is \textit{not} the shortest path. This might indicate that \texttt{hipMemcpyPeer} uses the bandwidth-maximizing path, instead of the shortest path, even for low transfer sizes. This is coherent with the purpose of \texttt{hipMemcpyPeer}, which allows large-size transfers, in contrast with granular accesses, e.g., performed with direct zero-copy access in unified memory.

\subsubsection{Bandwidth}
\label{sec:p2p-explicit-bw}

We measure the unidirectional bandwidth between each pair of GCDs using the p2pBandwidthLatencyTest benchmark, which relies on HIP APIs to perform copies. Figure~\ref{fig:p2p_bw_mat} presents the results. We can divide the results into two values of bandwidth: 50~GB/s and 37-38~GB/s. This result is not expected, as three distinct levels of bandwidth should be observed, namely 50, 100, 200~GB/s for single, dual, and quad link, respectively. In particular, the bandwidth measured for GCD pairs located on the same GPU (0-1, 2-3, 4-5 and 6-7) is on the order of 50~GB/s, which is significantly below the expected 200~GB/s bandwidth. This suggests that a single copy operation using \texttt{hipMemcpyPeer} cannot leverage the full bandwidth of an inter-GCD link. This behavior is documented by AMD, which indicates that the System Direct Memory Access (SDMA) engines, which are used for \texttt{hipMemcpy}, are tuned for PCIe-4.0 x16, and cannot utilize the full bandwidth of GPU-GPU Infinity Fabric interconnects. The advantage of using SDMA engine is that the use of \texttt{hipMemcpy} can be overlapped with computations, without affecting kernel performance. It is possible to disable the use of SDMA engines, to instead use a specialized ``blit'' copy kernel for \texttt{hipMemcpyPeer}, by setting the environment variable \texttt{HSA\_ENABLE\_PEER\_SDMA=0}. %

In addition, with these results, we can confirm that the path chosen for peer-to-peer communications with HIP memory copy API tends to optimize the bandwidth, and not the latency. Indeed, the 50 GB/s bandwidth measured for GCD pairs 1-7 and 3-5 can only be achieved through a three-hops path, longer than the shortest two-hops path for these pairs.

\begin{figure}[ht]
    \centering
    \includegraphics[width=\linewidth]{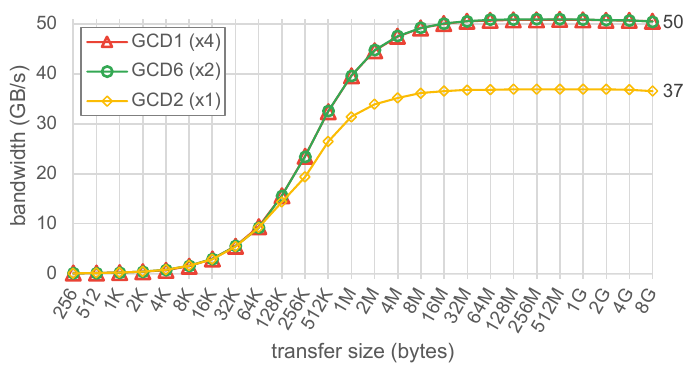}
    \caption{Peer-to-peer unidirectional bandwidth of a \texttt{hipMemcpyPeer} operation, from GCD0 to adjacent GCDs, measured with \commscope. The theoretical link bandwidth is indicated in parenthesis, as multiple of 50+50~GB/s links.}
    \label{fig:commscope_gcd0_gcdx}
\end{figure}

We further complete this high-level analysis by running the \commscope{ }benchmark for \texttt{hipMemcpyPeer}, performing the same peer-to-peer copy from GCD0 to the directly-connected GCDs, namely GCD\{1,2,6\}. Figure~\ref{fig:commscope_gcd0_gcdx} presents the bandwidth for data transfer sizes between 256~bytes to 8~GB. We reach comparable values as in Figure~\ref{fig:p2p_bw_mat}. The observation that the \texttt{hipMemcpyPeer} cannot utilize a full quad Infinity Fabric link remains valid in this case, for all transfer sizes. The bandwidth utilization for single, double, and quad Infinity Fabric links is 75\%, 50\% and 25\%, respectively.

%% file: 51_p2p_stream.tex
\subsection{Direct Memory Access}
\begin{figure}[ht]
    \centering
    \includegraphics[width=\linewidth]{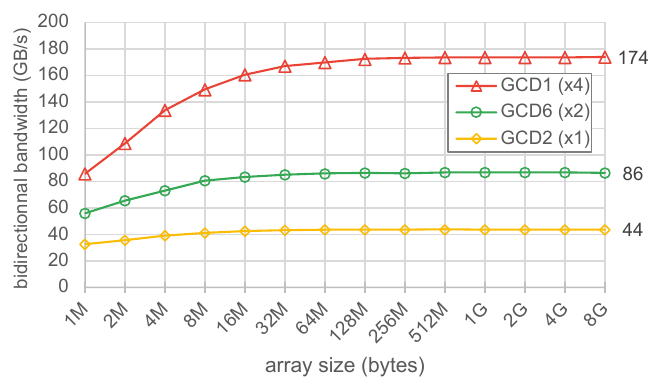}
    \caption{Bidirectional bandwidth measured with the STREAM copy kernel, executed on GCD0, with data placement on adjacent GCDs, namely GCD\{1,2,6\}, for increasing array sizes. The theoretical link bandwidth is indicated in parenthesis, as multiple of 50+50~GB/s links.}
    \label{fig:stream_p2p}
\end{figure}

\begin{figure}[ht]
    \centering
    \includegraphics[width=\linewidth]{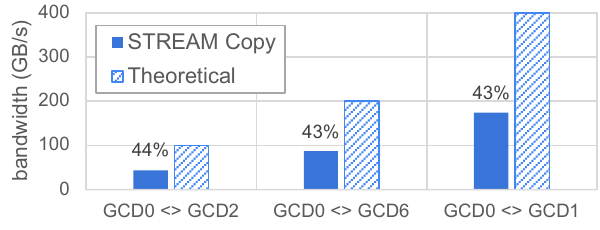}
    \caption{Peak bidirectional bandwidth, measured with the STREAM copy kernel, executed on GCD0, with data placement on adjacent GCDs, namely GCD\{1,2,6\}. Theoretical bandwidth is based on 50+50~GB/s for a simple Infinity Fabric link. Percentage values represent the ratio between measured and theoretical bandwidth.}
    \label{fig:stream_p2p_summary}
\end{figure}

To characterize performance of directly accessing peer-located memory, we use the STREAM copy kernel, in a similar fashion as for the CPU-GPU interconnect evaluation. The copy kernel is executed on GCD0, with data placed on adjacent GCDs, namely GCD\{1,2,6\}. As a reference, when using the same benchmark with data placement in local GCD0 memory, we observe a bandwidth of 1400~GB/s -- that is, 87\% of the theoretical 1.6~TB/s memory bandwidth. The copy bandwidth for the three placements is reported in Figure~\ref{fig:stream_p2p}, for increasing sizes, up to 8~GB. We observe three tiers of measured bandwidth values, representing the three tiers of Infinity Fabric links connecting GCD0 to its neighbors: single to GCD2, double to GCD6, and quad to GCD1. Figure~\ref{fig:stream_p2p_summary} presents the achieved bandwidth, along with the ratio of achieved theoretical bandwidth, based on 50+50~GB/s for a single Infinity Fabric link. For all placements, we observe that the achieved ratio of theoretical peak is 43-44\%. We do not observe the same bottleneck as identified when using \texttt{hipMemcpy} APIs, where using a quad Infinity Fabric link does not provide any improvement over using a dual link. As discussed previously, this is because kernel-level access to remote memory does not use the limited-bandwidth SDMA engines.

%% file: 53_p2p_MPI.tex
\subsection{GPU-aware MPI Communication}
MPI is commonly used in HPC applications for point-to-point communication. Recent GPU-aware MPI implementation poses as an alternative to HIP based communication. In this section, we use the vendor-provided Cray MPICH implementation, which supports direct peer-to-peer GPU communications. We use the OSU MPI point-to-point bandwidth benchmark~\cite{omb} that relies on the MPI\_ISend and MPI\_Recv to perform data movements between two MPI processes, each attached to one GPU. We found that the \texttt{HSA\_ENABLE\_SDMA} environment variable can affect the bandwidth, indicating that the data movements performed by the MPICH implementation may rely on a \texttt{hipMemcpy}-like interface.

\begin{figure}[ht]
    \centering
    \includegraphics[width=\linewidth]{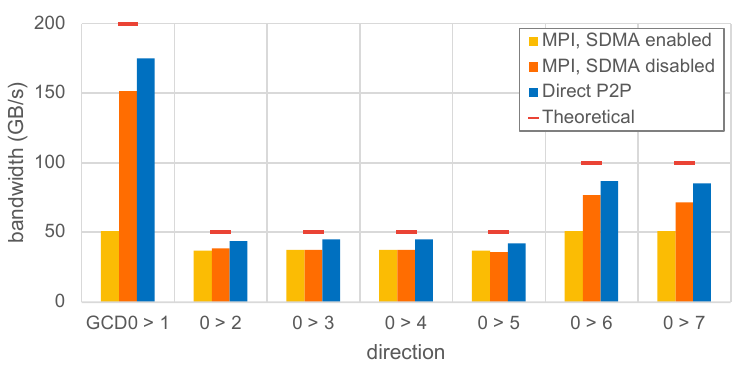}
    \caption{Unidirectional bandwidth of MPI point-to-point communication, measured with the OSU microbenchmarks (message size: 1~GiB), ``direct P2P'' is the bandwidth for a unidirectional STREAM copy from peer to local memory.}
    \label{fig:osu_p2p_bw}
\end{figure}

Figure~\ref{fig:osu_p2p_bw} shows the bandwidth for the OSU point-to-point bandwidth benchmark that sends data from GCD0 to other GCDs. Results for both MPI, and direct peer-to-peer communication using a STREAM-like kernel are presented. For the MPI benchmark, we provide the results using SDMA engines (SDMA enabled and \texttt{hipMemcpy}-like) and using direct copy kernel (SDMA disabled and copy kernel). 

As expected, the use of SDMA provides a sub-optimal bandwidth, below 50~GB/s, similarly as in the peer-to-peer results for explicit data movements (in Section~\ref{sec:p2p-explicit-bw}). Note that the use of SDMA engine has the advantage that the MPI\_ISend operation can be overlapped with GPU kernel execution. As the maximum unidirectional bandwidth between GCD0 and GCD\{2,3,4,5\} is 50~GB/s, the use of SDMA engines still provides a high utilization of the available bandwidth. However, results from GCD0 to GCD\{1,6,7\} are different, as those links exhibit a higher available bandwidth. Here, the SDMA-enabled MPI transfer only reaches 50~GB/s -- below 50\% for a dual Infinity Fabric link, and 25\% for a quad link. Therefore, if no overlap of data transfer with GPU kernel execution is required or possible, it is advised to disable SDMA data transfer, by setting the environment variable \texttt{HSA\_ENABLE\_SDMA=0}.

Interestingly, the SDMA-disabled MPI transfer exhibits a 10-15\% lower bandwidth than the direct peer-to-peer copy kernel. This difference could come from the overhead in MPI communications, compared to the direct implementation of a copy kernel in HIP. Furthermore, we observe that transferring data from GCD0 to a non-neighbor GCD, namely GCD{3,4,5,7}, does not exhibit significant difference in measured bandwidth compared to neighbor GCDs. 

%% file: 54_collectives.tex
\section{GPU Collective Communication}
When multiple GPUs need to communicate, the use of collective communication can be more efficient than purposefully-designed algorithm which leverage point-to-point communication. In this section, we investigate the latency of five commonly-used collectives -- Reduce, Broadcast, AllReduce, ReduceScatter, and AllGather. These collectives can be categorized into two types: for Reduce and Broadcast only one communication pass from all GCDs to one GCD is required (or from one GCD to all others). For AllReduce, AllGather, and ReduceScatter, two communication passes are required, where the first pass aggregates data from all GCDs into one result and then disseminates the result back to all GCDs. We use the collective OSU micro-benchmarks and investigate two common interfaces: MPI collectives and RCCL collectives.

\begin{figure*}
    \begin{subfigure}{0.195\linewidth}
        \centering
        \includegraphics[width=\linewidth]{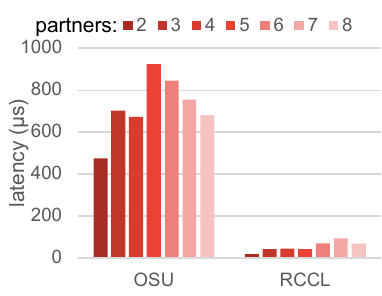}
        \caption{Reduce}
        \label{fig:lat_reduce}
    \end{subfigure} %
    \begin{subfigure}{0.195\linewidth}
        \centering
        \includegraphics[width=\linewidth]{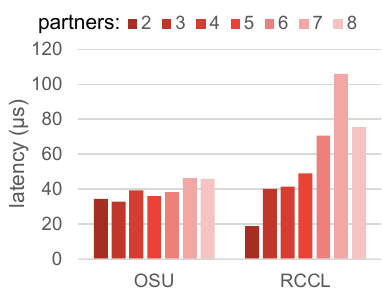}
        \caption{Broadcast}
    \end{subfigure}
    \begin{subfigure}{0.195\linewidth}
        \centering
        \includegraphics[width=\linewidth]{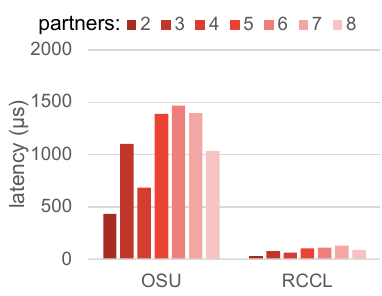}
        \caption{AllReduce}
    \end{subfigure} %
    \begin{subfigure}{0.195\linewidth}
        \centering
        \includegraphics[width=\linewidth]{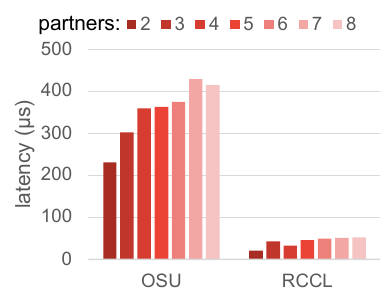}
        \caption{ReduceScatter}
    \end{subfigure}
    \centering
    \begin{subfigure}{0.195\linewidth}
        \centering
        \includegraphics[width=\linewidth]{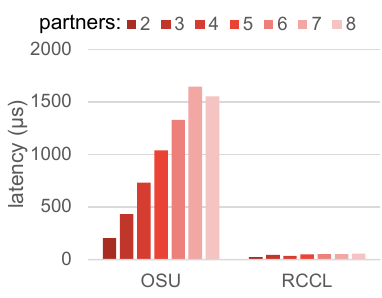}
        \caption{AllGather}
        \label{fig:lat_allgather}
    \end{subfigure}
    \caption{Latency of five collective operations (\subref{fig:lat_reduce}-\subref{fig:lat_allgather}) with OSU micro-benchmarks, compared to RCCL, using two to eight communication partners (message size: 1 MiB), each partner uses one GPU. For MPI, a partner is a MPI process, for RCCL, a partner is a CPU thread.}
    \label{fig:collectives}
\end{figure*}
\begin{figure}[bt]
    \centering
    \includegraphics[width=\linewidth]{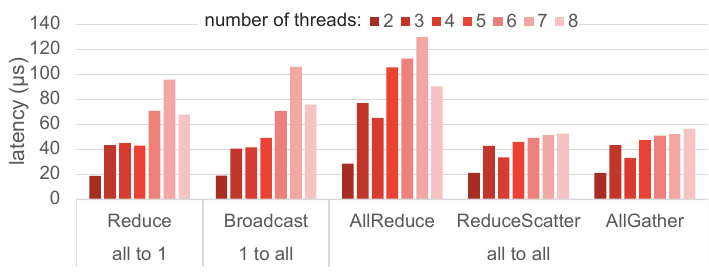}
    \caption{Latency of five collective operations in RCCL, with two to eight CPU threads. Each thread takes part in the collective operation, one GPU per thread.}
    \label{fig:lat_rccl}
\end{figure}

Analytically, we can approximate the lower bound of latency of these two categories of collective communication from the latency matrix presented in Figure~\ref{fig:p2p_lat_mat}, which reports the latency for communication within all pairs of GCDs. Taking the lowest GCD-GCD latency of 8.7~$\mu$s, single-round collectives latency has a lower bound of 8.7~$\mu$s, and dual-round collectives should have a latency of at least 17.4~$\mu$s.

The measured latency in RCCL collectives is presented in Figure~\ref{fig:lat_rccl}. For two threads, the lowest measured latency for all-to-all collectives is close to the lowest bound of 17.4~$\mu$s. When increasing the number of threads above 2, the latency increases as expected, as the implementation of all-to-all operations may not follow a simple pattern that can achieve the lowest latency bound. Interestingly, for Reduce, Broadcast, and AllReduce collectives, the latency drops when increasing from 7 to 8 threads, possibly due to the more balanced communication pattern when all eight GPUs are used.

Figure~\ref{fig:collectives} compares the measured latency for MPI and RCCL collectives on a 1~MB array. Our evaluation results show that RCCL is more efficient than MPI collectives for all tested collectives, except for broadcast. A similar finding is also reported in previous work~\cite{chen2023mpi}. The performance overhead on MPI collectives could come from memory mapping overhead, where extra overhead is needed to exchange and map HIP pointers into each process' virtual memory space to support CPU-side inter-process communication in MPI. Note that although RCCL is more efficient within a single node compared to MPI, RCCL relies on MPI for multi-node communication, and thus, the conclusion for multi-node collectives can be different.

%% file: 60_related.tex
\section{Related Works}
\label{sec:related}
Various aspects of the AMD MI250X GPU have been investigated before. Pearson et al.~\cite{pearson2023interconnect} focus on the interconnect performance across on-node MI250X GPUs. Schieffer et al.~\cite{schieffer2024rise} focus on the matrix core units on AMD MI250X GPUs. In contrast, our work focuses on the efficiency of various programming interfaces for data access and communication on multi-GPU nodes. Leinhauser et al.~\cite{leinhauser2022metrics} design and develop an instruction roofline model for AMD GPUs. They focus on the effect of problem size and GPU launch configurations on roofline performance for V100, A100, MI100, and MI250X graphics processing units. Eberius et al.~\cite{eberius2022understanding} extend the roofline model to consider problem size and characterize strong scaling on AMD250X GPUs using saturation problem sizes as an additional performance metric. Punniyamurthy et al.~\cite{punniyamurthy2023gpu} investigate the advantages of fusing computation with dependent collective communication by exploiting GPU-initiated communication and communication across different GPUs in Machine Learning workloads.

Other works have also evaluated the performance of programming systems and libraries for intra-node communication. Godoy et al.~\cite{godoy2023evaluating} evaluate the performance and portability of high-level programming models, including Julia, Python/Numba, and Kokkos on HPC nodes with multiple AMD GPUs. The HipBone~\cite{chalmers2023hipbone} proxy app of Nek5000, is developed as a performance-portable GPU C++ version to characterize the performance of different GPUs, including the AMD MI250X.

On AMD GPU memory systems, Jin et al.~\cite{jin2022evaluating} focus on the performance of unified memory using the HIP programming interface. They conclude that while unified memory can improve programmability, its usage comes with a significant overhead impacting performance on AMD GPUs. Similar work has been carried out on Nvidia GPUs. Chien et al.~\cite{chien2019performance} study the impact of memory policies and hints on the CUDA managed memory on Nvidia GPUs. Schieffer et al.~\cite{schieffer2024harnessing} study the integrated CPU-GPU system memory, available on the Nvidia Grace Hopper Superchip. Li et al.~\cite{li2018tartan} analyze the interconnect of Nvidia GPUs through the Tartan benchmark suite. In this work, similar experiments have been performed on AMD GPUs allowing for a comparison between these two architectures. Several application-specific works on multi-GPU also investigate performance on AMD GPUs. They focus on improved memory management and data transfers, such as multi-GPU quantum computing simulations ~\cite{horii2023efficient,faj2023quantum}, and graph-processing workloads~\cite{min2020emogi}.

%% file: 70_conclusion.tex
\section{Conclusion}
\label{sec:conclusion}
In this work, we evaluated the various data movement options on multi-GPU nodes that use Infinity Fabric to interconnect a CPU with eight AMD Instinct GCDs. The testbed represents a topology similar to that of the Frontier supercomputer, the first exascale supercomputer. Our testing methodology started with identifying the peak hardware capacity and evaluates various software options for data movements, including CPU-GPU, point-to-point GPU-GPU, and GPU collectives. Our results quantified the impact of memory allocation strategies on data movement between AMD CPU and GPUs. For the performance of collective communication, we compared MPI and RCCL libraries on an AMD multi-GPU node. Our results highlights that the complex nature of the multi-GPU topology must be taken into account to achieve high utilization of hardware capabilities in data movement, despite being abstracted into a simple yet flexible programming model. In particular, attention must be focused on environment configuration, task-to-GPU mapping, and choice of interface and libraries.

%% file: main.bbl
\begin{thebibliography}{10}
\providecommand{\url}[1]{#1}
\csname url@samestyle\endcsname
\providecommand{\newblock}{\relax}
\providecommand{\bibinfo}[2]{#2}
\providecommand{\BIBentrySTDinterwordspacing}{\spaceskip=0pt\relax}
\providecommand{\BIBentryALTinterwordstretchfactor}{4}
\providecommand{\BIBentryALTinterwordspacing}{\spaceskip=\fontdimen2\font plus
\BIBentryALTinterwordstretchfactor\fontdimen3\font minus \fontdimen4\font\relax}
\providecommand{\BIBforeignlanguage}[2]{{%
\expandafter\ifx\csname l@#1\endcsname\relax
\typeout{** WARNING: IEEEtran.bst: No hyphenation pattern has been}%
\typeout{** loaded for the language `#1'. Using the pattern for}%
\typeout{** the default language instead.}%
\else
\language=\csname l@#1\endcsname
\fi
#2}}
\providecommand{\BIBdecl}{\relax}
\BIBdecl

\bibitem{frontier-user}
\BIBentryALTinterwordspacing
OLCF, ``Frontier user guide,'' 2023. [Online]. Available: \url{https://docs.olcf.ornl.gov/systems/frontier_user_guide.html}
\BIBentrySTDinterwordspacing

\bibitem{geenen2024digital}
T.~Geenen, N.~Wedi, S.~Milinski, I.~Hadade, B.~Reuter, S.~Smart, J.~Hawkes, E.~Kuwertz, T.~Quintino, E.~Danovaro \emph{et~al.}, ``Digital twins, the journey of an operational weather system into the heart of destination earth,'' \emph{Procedia Computer Science}, vol. 240, pp. 99--108, 2024.

\bibitem{karp2023large}
M.~Karp, D.~Massaro, N.~Jansson, A.~Hart, J.~Wahlgren, P.~Schlatter, and S.~Markidis, ``Large-scale direct numerical simulations of turbulence using gpus and modern fortran,'' \emph{The International Journal of High Performance Computing Applications}, vol.~37, no.~5, pp. 487--502, 2023.

\bibitem{jansson2023exploring}
N.~Jansson, M.~Karp, A.~Perez, T.~Mukha, Y.~Ju, J.~Liu, S.~P{\'a}ll, E.~Laure, T.~Weinkauf, J.~Schumacher \emph{et~al.}, ``Exploring the ultimate regime of turbulent rayleigh--b{\'e}nard convection through unprecedented spectral-element simulations,'' in \emph{Proceedings of the International Conference for High Performance Computing, Networking, Storage and Analysis}, 2023, pp. 1--9.

\bibitem{andersson2022breaking}
M.~I. Andersson, N.~A. Murugan, A.~Podobas, and S.~Markidis, ``Breaking down the parallel performance of gromacs, a high-performance molecular dynamics software,'' in \emph{International Conference on Parallel Processing and Applied Mathematics}.\hskip 1em plus 0.5em minus 0.4em\relax Springer, 2022, pp. 333--345.

\bibitem{williams2024understanding}
J.~J. Williams, A.~Bhole, D.~Kierans, M.~Hoelzl, I.~Holod, W.~Tang, D.~Tskhakaya, S.~Costea, L.~Kos, A.~Podolnik \emph{et~al.}, ``Understanding large-scale plasma simulation challenges for fusion energy on supercomputers,'' \emph{arXiv preprint arXiv:2407.00394}, 2024.

\bibitem{markidis2023enabling}
S.~Markidis, ``Enabling quantum computer simulations on amd gpus: a hip backend for google's qsim,'' in \emph{Proceedings of the SC'23 Workshops of The International Conference on High Performance Computing, Network, Storage, and Analysis}, 2023, pp. 1478--1486.

\bibitem{loh2023research}
G.~H. Loh, M.~J. Schulte, M.~Ignatowski, V.~Adhinarayanan, S.~Aga, D.~Aguren, V.~Agrawal, A.~M. Aji, J.~Alsop, P.~Bauman \emph{et~al.}, ``A research retrospective on amd's exascale computing journey,'' in \emph{Proceedings of the 50th Annual International Symposium on Computer Architecture}, 2023, pp. 1--14.

\bibitem{frontier}
\BIBentryALTinterwordspacing
OLCF, ``Frontier supercomputer debuts as world’s fastest, breaking exascale barrier,'' 2023. [Online]. Available: \url{https://www.ornl.gov/news/frontier-supercomputer-debuts-worlds-fastest-breaking-exascale-barrier}
\BIBentrySTDinterwordspacing

\bibitem{lumi}
\BIBentryALTinterwordspacing
{LUMI consortium}, ``Lumi’s full system architecture revealed,'' 2021. [Online]. Available: \url{https://www.lumi-supercomputer.eu/lumis-full-system-architecture-revealed/}
\BIBentrySTDinterwordspacing

\bibitem{mi250_microarch}
\BIBentryALTinterwordspacing
AMD, ``Amd instinct mi250 microarchitecture,'' 2024. [Online]. Available: \url{https://rocm.docs.amd.com/en/latest/conceptual/gpu-arch/mi250.html}
\BIBentrySTDinterwordspacing

\bibitem{pearson2019}
\BIBentryALTinterwordspacing
C.~Pearson, A.~Dakkak, S.~Hashash, C.~Li, I.-H. Chung, J.~Xiong, and W.-M. Hwu, ``Evaluating characteristics of cuda communication primitives on high-bandwidth interconnects,'' in \emph{Proceedings of the 2019 ACM/SPEC International Conference on Performance Engineering}, ser. ICPE '19.\hskip 1em plus 0.5em minus 0.4em\relax New York, NY, USA: Association for Computing Machinery, 2019, p. 209–218. [Online]. Available: \url{https://doi.org/10.1145/3297663.3310299}
\BIBentrySTDinterwordspacing

\bibitem{cudasamples}
\BIBentryALTinterwordspacing
Nvidia, ``Cuda samples,'' 2018. [Online]. Available: \url{https://github.com/NVIDIA/cuda-samples}
\BIBentrySTDinterwordspacing

\bibitem{omb}
\BIBentryALTinterwordspacing
``Osu micro-benchmarks,'' 2001. [Online]. Available: \url{http://mvapich.cse.ohio-state.edu/benchmarks/}
\BIBentrySTDinterwordspacing

\bibitem{pearson2023interconnect}
C.~Pearson, ``Interconnect bandwidth heterogeneity on amd mi250x and infinity fabric,'' \emph{arXiv preprint arXiv:2302.14827}, 2023.

\bibitem{chen2023mpi}
C.-C. Chen, K.~Shafie~Khorassani, P.~Kousha, Q.~Zhou, J.~Yao, H.~Subramoni, and D.~K. Panda, ``Mpi-xccl: A portable mpi library over collective communication libraries for various accelerators,'' in \emph{Proceedings of the SC'23 Workshops of The International Conference on High Performance Computing, Network, Storage, and Analysis}, 2023, pp. 847--854.

\bibitem{schieffer2024rise}
G.~Schieffer, D.~A. De~Medeiros, J.~Faj, A.~Marathe, and I.~Peng, ``On the rise of amd matrix cores: Performance, power efficiency, and programmability,'' in \emph{2024 IEEE International Symposium on Performance Analysis of Systems and Software (ISPASS)}.\hskip 1em plus 0.5em minus 0.4em\relax IEEE, 2024, pp. 132--143.

\bibitem{leinhauser2022metrics}
M.~Leinhauser, R.~Widera, S.~Bastrakov, A.~Debus, M.~Bussmann, and S.~Chandrasekaran, ``Metrics and design of an instruction roofline model for amd gpus,'' \emph{ACM Transactions on Parallel Computing}, vol.~9, no.~1, pp. 1--14, 2022.

\bibitem{eberius2022understanding}
D.~Eberius, P.~Roth, and D.~M. Rogers, ``Understanding strong scaling on gpus using empirical performance saturation size,'' in \emph{2022 IEEE/ACM International Workshop on Performance, Portability and Productivity in HPC (P3HPC)}.\hskip 1em plus 0.5em minus 0.4em\relax IEEE, 2022, pp. 26--35.

\bibitem{punniyamurthy2023gpu}
K.~Punniyamurthy, B.~M. Beckmann, and K.~Hamidouche, ``Gpu-initiated fine-grained overlap of collective communication with computation,'' \emph{arXiv preprint arXiv:2305.06942}, 2023.

\bibitem{godoy2023evaluating}
W.~F. Godoy, P.~Valero-Lara, T.~E. Dettling, C.~Trefftz, I.~Jorquera, T.~Sheehy, R.~G. Miller, M.~Gonzalez-Tallada, J.~S. Vetter, and V.~Churavy, ``Evaluating performance and portability of high-level programming models: Julia, python/numba, and kokkos on exascale nodes,'' \emph{arXiv preprint arXiv:2303.06195}, 2023.

\bibitem{chalmers2023hipbone}
N.~Chalmers, A.~Mishra, D.~McDougall, and T.~Warburton, ``Hipbone: A performance-portable graphics processing unit-accelerated c++ version of the nekbone benchmark,'' \emph{The International Journal of High Performance Computing Applications}, p. 10943420231178552, 2023.

\bibitem{jin2022evaluating}
Z.~Jin and J.~S. Vetter, ``Evaluating unified memory performance in hip,'' in \emph{2022 IEEE International Parallel and Distributed Processing Symposium Workshops (IPDPSW)}.\hskip 1em plus 0.5em minus 0.4em\relax IEEE, 2022, pp. 562--568.

\bibitem{chien2019performance}
S.~Chien, I.~Peng, and S.~Markidis, ``Performance evaluation of advanced features in cuda unified memory,'' in \emph{2019 IEEE/ACM Workshop on Memory Centric High Performance Computing (MCHPC)}.\hskip 1em plus 0.5em minus 0.4em\relax IEEE, 2019, pp. 50--57.

\bibitem{schieffer2024harnessing}
G.~Schieffer, J.~Wahlgren, J.~Ren, J.~Faj, and I.~Peng, ``Harnessing integrated cpu-gpu system memory for hpc: a first look into grace hopper,'' in \emph{Proceedings of the 53rd International Conference on Parallel Processing}, 2024, pp. 199--209.

\bibitem{li2018tartan}
A.~Li, S.~L. Song, J.~Chen, X.~Liu, N.~Tallent, and K.~Barker, ``Tartan: evaluating modern gpu interconnect via a multi-gpu benchmark suite,'' in \emph{2018 IEEE International Symposium on Workload Characterization (IISWC)}.\hskip 1em plus 0.5em minus 0.4em\relax IEEE, 2018, pp. 191--202.

\bibitem{horii2023efficient}
H.~Horii, C.~Wood \emph{et~al.}, ``Efficient techniques to gpu accelerations of multi-shot quantum computing simulations,'' \emph{arXiv preprint arXiv:2308.03399}, 2023.

\bibitem{faj2023quantum}
J.~Faj, I.~Peng, J.~Wahlgren, and S.~Markidis, ``Quantum computer simulations at warp speed: Assessing the impact of gpu acceleration: A case study with ibm qiskit aer, nvidia thrust \& cuquantum,'' in \emph{2023 IEEE 19th International Conference on e-Science (e-Science)}.\hskip 1em plus 0.5em minus 0.4em\relax IEEE, 2023, pp. 1--10.

\bibitem{min2020emogi}
S.~W. Min, V.~S. Mailthody, Z.~Qureshi, J.~Xiong, E.~Ebrahimi, and W.-m. Hwu, ``Emogi: Efficient memory-access for out-of-memory graph-traversal in gpus,'' \emph{arXiv preprint arXiv:2006.06890}, 2020.

\end{thebibliography}
